\documentclass[pss]{wiley2sp} 

\usepackage{amsmath} 
\usepackage{graphicx}
\usepackage{rotating}  
\usepackage{varwidth,array} 
\usepackage{widetext} 

\def\ep{\varepsilon}  
\newcommand{\ds}{\displaystyle}
\newcolumntype{L}{>{\varwidth[c]{\linewidth}}l<{\endvarwidth}} 

\begin{document}
\hyphenation{ani-so-tro-py re-fra-cti-vity qua-dra-tic mag-ne-to trans-port mag-ne-to-trans-port}

\title{Quadratic-in-magnetization permittivity and conductivity tensor in cubic crystals}  
\titlerunning{Quadratic-in-magnetization permittivity and \ldots}
\author{Jana Hamrlov\'a, Jaroslav Hamrle, Kamil Postava and Jarom{\'\i}r Pi\v{s}tora}
\authorrunning{J. Hamrlov\'a et al.}
\mail{e-mail \textsf{jana.hamrlova@vsb.cz}}

\institute{Nanotechnology Centre, VSB - Technical University of Ostrava, 17. listopadu 15, 708 33 Ostrava-Poruba, Czech Republic}

\abstract{%
We present diagonal and off-diagonal elements of the permittivity and conductivity tensor up to the second order in magnetization for cubic crystals. We express all tensor elements as a function of a general sample orientation, for arbitrary magnetization direction and for (001), (011) and (111) surface orientations. Finally, we discuss, how to extract values of quadratic elements $G$ of the second order permittivity tensors for different sample surface orientations from both experiment and ab-initio calculations. 
} 

\received{XXXX, revised XXXX, accepted XXXX} 
\published{XXXX} 


\maketitle

\section{Introduction}
  
There is a vast number of physical phenomena, which scale with quadratic form of magnetization (also called effects quadratic in magnetization). Within the transport community the well-known examples are  
anisotropy magnetoresistance (AMR) \cite{mallinson,doring1938} or
longitudinal Hall effect \cite{grebner60}.
Optical effects quadratic in magnetization include quadratic magneto-optical Kerr effect (QMOKE) \cite{vis86,pos02,ham07quad}, magnetic linear dichroism (MLD) and birefringence \cite{usachev2005} (used also to investigate antiferromagnets \cite{ferre1983}), magneto-refractivity \cite{hrabovsky2009,kravets2002}. For overview of optical quadratic effects, see reviews \cite{pisarev1975,ferre1984,ferre2001}.
Recently, quadratic effects are intensively studied within X-ray optical range, measuring X-ray magnetic linear dichroism (XMLD) \cite{mertins01,valencia10}.

Although all those effects are not usually understand as a single phenomena, basically all of them originate from non-zero conductivity $\sigma_{ij}$ (or permittivity $\ep_{ij}$) tensor, defines as $j_i=\sigma_{ijkl}E_jM_kM_l$, where $j_i$ is the vector of current density, $\sigma_{ijkl}$ is the fourth-rank tensor, being quadratic (second order) in magnetization (components $M_k$, $M_l$) and $E_j$ is vector of the electric field. 
The difference between different effects mentioned above is than basically (i) photon energy range and (ii) whether the measured quantity is related with the diagonal or the off-diagonal part of the conductivity or permittivity tensor. 
%
%
Table~\ref{t:tech} overviews techniques studying the off-diagonal permittivity or conductivity tensor in the first and in the second order in magnetization, and in the difference of the diagonal terms. 
\begin{table*}
\begin{tabular}{|m{0.31\textwidth}||>{\centering}m{0.13\textwidth}|>{\centering}m{0.26\textwidth}|>{\centering}m{0.19\textwidth}|}
\hline
&conductivity & magneto-optics & X-ray
\tabularnewline\hline\hline
 off-diagonals linear in $\vec{M}$, $\varepsilon_{ij}^{(1)}$, $i\neq j$  
 & Hall effect
 & MOKE \\ magnetic circular dichroism and birefringence (MCD, MCB)
 & X-ray  magnetic circular dichroism (XMCD)
\tabularnewline\hline
 off-diagonals quadratic in $\vec{M}$ $\varepsilon_{ij}^{(2)}$, $i\neq j$ 
 & longitudinal Hall (quadratic Hall$^*$) 
 & QMOKE  
 & quadratic XMCD$^*$  
\tabularnewline\hline
 difference of diagonals, $\varepsilon^{(2)}_{ii}-\varepsilon^{(2)}_{jj}$ ($i\neq j$)
 \newline being quadratic in $\vec{M}$
 &
 AMR & Voigt effect (Cotton-Mouton)\\ magnetic linear dichroism (MLD)\\  Sch\"afer-Hubert effect & X-ray magnetic dischroism (XMLD)
\tabularnewline\hline
\end{tabular}
\caption{Techniques used to study first and second order effects in magnetization. Techniques denoted by $^*$ are not experimentally demonstrated yet. XMCD stands for X-ray magnetic circular dichroism.}
\label{t:tech}
\end{table*}

With some generalization, basically every ma\-gne\-to-trans\-port phenomena is expected to have its second-order counterpart. Hence, existence of effects such as quadratic spin-Hall effect or quadratic magneto-Seebeck effect can be envisaged, although those effects have not been demonstrated experimentally neither by ab-initio calculations yet.

Within this Article, based on symmetry arguments, we determine the diagonal and the off-diagonal elements of the permittivity tensor in cubic crystal up to the second order in magnetization. We express all tensor elements for arbitrary magnetization direction and for (001), (011) and (111) surface orientations of the sample. We also distinguish two different kinds of symmetries in cubic crystals, (i) the crystals with the symmetry classes 23 and $m3$ and (ii) after adding the symmetry operation the rotation about $180^\circ$ about $xy$ axis, the cubic crystals with the symmetry classes  $\overline{4}3m$, 432 and $m3m$ are described. 

\section{Motivation}

Within the complex representation of the Maxwell equations, the complex (relative) permittivity tensor $\varepsilon_{ij}$ and the  complex conductivity tensor $\sigma_{ij}$ describe the same physical phenomena, with conversion between them being (in SI) $\varepsilon_{ij}=\delta_{ij}+i \sigma_{ij}/\omega$, where $\delta_{ij}$ is Kronecker delta and $\omega$ is the photon frequency (here, $\sigma_{ij}$ is expressed in units [s$^{-1}$]). 
Hence, from symmetry point of view, the symmetry form of $\varepsilon_{ij}$ and $\sigma_{ij}$ are exactly the same. Therefore, in following we express tensors' symmetry form solely for permittivity. The form of conductivity tensor is obtained simply by interchanging variables $\sigma_{ij}\leftrightarrow\varepsilon_{ij}$. 
As we express the tensors' symmetry form in permittivity $\varepsilon_{ij}$, we later call this tensor by its historical name as magneto-optical permittivity tensor, although, as told above, the presented symmetry analysis are equal for conductivity investigations.

Those symmetry arguments analysis are important in following cases: 

(i) Experimental evidence of the second order effects may be difficult, as those effects can be small or they can be mimiced by artefacts of the experiments. Therefore, to confirm that the measured effect is a quadratic effect, one should measure interplay of magnetic field direction, crystallography axis direction and experimental geometry. For example, when one founds geometry where the first order effects vanishes and when the behaviour of measured effects agrees with the form predicted by the symmetry arguments, then the presence of quadratic effects is clearly proven \cite{pos02,ham07quad}. Therefore, detailed understanding of symmetry of quadratic effects helps to optimize measurement geometry for each experimental technique.

(ii) Furthermore, those combined measurements allow to determine quadratic elements of the second order permittivity tensors (in following denoted as $G_{44}$, $\Delta G$, $\Delta\Gamma$). The symmetry arguments are important to determine experimental geometry, in which those elements can be extracted. 
This opens new spectroscopy branch where spectra of $G_{44}$, $\Delta G$, $\Delta\Gamma$ quadratic elements are investigated. Up to now, no spectra of the quadratic elements are available and the complex values of $G_{44}$, $\Delta G$ were determined only for Fe(001) for a single wavelength \cite{buchmeier2002}.

(iii) The symmetry arguments presented here are also important for ab-initio calculations, to verify correctness of the calculated permittivity or conductivity on applied direction of the magnetization as well as to express ab-initio spectra of $G_{44}$, $\Delta G$, $\Delta\Gamma$ quadratic elements. 
Nowadays ab-initio codes calculate total permittivity or conductivity, as a function of the applied direction of the magnetic field. Hence, in order to separate the quadratic elements from the ab-initio-calculated total permittivity or conductivity, symmetry arguments predictions of the dependence of the tensors on magnetization direction must be used and fitted to the ab-initio values of the tensor elements.

(iv) Last, note that definition of second order permittivity elements and their dependence on magnetic field and sample orientation contain several sign conventions. However, to correctly compare extracted second order permittivity tensors from both experiment and ab-initio calculations, the sign conventions must be handled carefully and correctly. The tensor transformation presented within this Article were done with a careful definition of all sign conventions.
Therefore, this Article may also establish the sign conventions for second order effects. This will be particularly important in expected spectroscopies of the second order permittivity tensors.

\section{Magneto-optical permittivity tensor}

In magneto-optically active materials we can evolute permittivity tensor with respect to the components of magnetization ${\mathbf M}$
acting on the material \cite{vis86}
\begin{equation}
\begin{split}
\varepsilon_{ij} =& \varepsilon_{ij}^{(0)}+\left[\frac{\partial \varepsilon_{ij}}{\partial M_{k}}\right]_{{\mathbf M}=0}M_k
\\ 
&+\frac{1}{2}\left[\frac{\partial^2 \varepsilon_{ij}}{\partial M_k\partial M_l}\right]_{{\mathbf M}=0}M_kM_l+\ldots \\
 = & \varepsilon_{ij}^{(0)}+K_{ijk}M_k+G_{ijkl}M_kM_l+\ldots 
\end{split}
\end{equation}
where $\varepsilon_{ij}^{(0)}=\varepsilon_{ji}^{(0)}$ are components of permittivity tensor of material without acting magnetization ${\mathbf M}$. Contribution to the permittivity tensor $\varepsilon_{ij}^{(1)}=K_{ijk}M_k$ denotes contribution linear in magnetization ${\mathbf M}$ where a third rank tensor $K_{ijk}$ is called {\it linear magneto-optical tensor}.
Contribution to the permittivity tensor $\varepsilon_{ij}^{(2)}=G_{ijkl}M_kM_l$ is quadratic in magnetization ${\mathbf M}$ and is described by the fourth rank 
{\it quadratic magneto-optical tensor} $G_{ijkl}$.


Magneto-optical permittivity tensor must fulfil Onsager relation
\begin{equation}
\varepsilon_{ij}({\mathbf M})=\varepsilon_{ji}(-{\mathbf M})\,.
\end{equation}
Hence, the quadratic magneto-optical contribution to the permittivity tensor $\varepsilon_{ij}^{(2)}$ must be symmetric \cite{vis86},
$\varepsilon_{ij}^{(2)} = \varepsilon_{ji}^{(2)}$,  
which implies
$G_{ijkl}  =  G_{jikl} = G_{ijlk} = G_{jilk}$. Thus, the quadratic magneto-optical permittivity tensor can be rewritten in simplified form \cite{vis86}
\begin{equation}
\label{eq:qgen}
\left[ 
\begin{array}{c}
\varepsilon_{11}^{(2)}\\
\varepsilon_{22}^{(2)}\\
\varepsilon_{33}^{(2)}\\
\varepsilon_{23}^{(2)}\\
\varepsilon_{31}^{(2)}\\
\varepsilon_{12}^{(2)}
\end{array}
\right] = \left[ 
\begin{array}{cccccc}
G_{1111} & G_{1122} & G_{1133} & 2G_{1123} & 2G_{1131} & 2G_{1112} \\
G_{2211} & G_{2222} & G_{2233} & 2G_{2223} & 2G_{2231} & 2G_{2212} \\
G_{3311} & G_{3322} & G_{3333} & 2G_{3323} & 2G_{3331} & 2G_{3312} \\
G_{2311} & G_{2322} & G_{2333} & 2G_{2323} & 2G_{2331} & 2G_{2312} \\
G_{3111} & G_{3122} & G_{3133} & 2G_{3123} & 2G_{3131} & 2G_{3112} \\
G_{1211} & G_{1222} & G_{1233} & 2G_{1223} & 2G_{1231} & 2G_{1212} 
\end{array}
\right] \left[ 
\begin{array}{c}
M_1^2 \\
M_2^2 \\
M_3^2\\
M_2M_3\\
M_3M_1\\
M_1M_2
\end{array}
\right]
\end{equation}
being a general expression of the second order magneto-optical permittivity tensor.

\subsection{Quadratic magneto-optical tensor in cubic crystals}

\begin{figure}[b]
\centering
\includegraphics[width=.15\textwidth,angle=0]{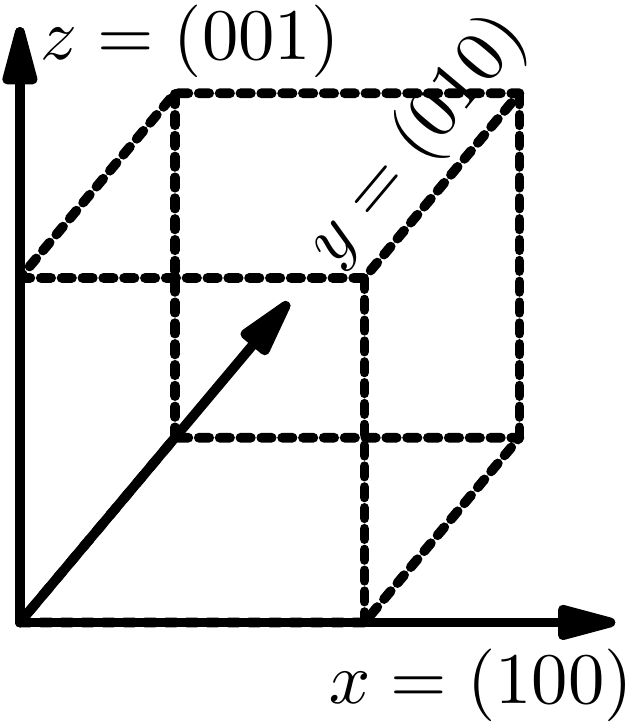}
\caption{\label{f:one}%
Principal axes of cubic crystal.}
\end{figure}

In our work we are focusing solely on cubic crystals. 
Let's suppose that coordinate system corresponding to the principal axes of the crystal (Fig.~\ref{f:one}).
Then (for the crystal classes 23 and $m3$), the quadratic magneto-optical tensor in cubic crystals in general form is \cite{bhagavantam}
\begin{equation}
\mathbf{G}=
 \left[ 
\begin{array}{cccccc}
G_{11} & G_{12} & G_{21} & 0 & 0 & 0 \\
G_{21} & G_{11} & G_{12} & 0 & 0 & 0 \\
G_{12} & G_{21} & G_{11} & 0 & 0 & 0 \\
0 & 0 & 0 & 2G_{44} & 0 & 0 \\
0 & 0 & 0 & 0 & 2G_{44} & 0 \\
0 & 0 & 0 & 0 & 0 & 2G_{44} 
\end{array}
\right],
\label{eq:kubG}
\end{equation}
where short index notation of $G_{ijkl}$ was used. 
Adding the symmetry operation (the rotation about $180^\circ$ about $xy$ axis) we have for cubic crystals with the crystal classes  $\overline{4}3m$, 432
and $m3m$ \cite{bhagavantam}
\begin{equation}
\mathbf{G}=
 \left[ 
\begin{array}{cccccc}
G_{11} & G_{12} & G_{12} & 0 & 0 & 0 \\
G_{12} & G_{11} & G_{12} & 0 & 0 & 0 \\
G_{12} & G_{12} & G_{11} & 0 & 0 & 0 \\
0 & 0 & 0 & 2G_{44} & 0 & 0 \\
0 & 0 & 0 & 0 & 2G_{44} & 0 \\
0 & 0 & 0 & 0 & 0 & 2G_{44} 
\end{array}
\right].
\label{eq:kubGG}
\end{equation}
Adding further symmetry operation then describes isotropic medium \cite{bhagavantam}
\begin{equation}
 \mathbf{G}=\left[ 
\begin{array}{cccccc}
G_{11} & G_{12} & G_{12} & 0 & 0 & 0 \\
G_{12} & G_{11} & G_{12} & 0 & 0 & 0 \\
G_{12} & G_{12} & G_{11} & 0 & 0 & 0 \\
0 & 0 & 0 & G_{11}-G_{12} & 0 & 0 \\
0 & 0 & 0 & 0 & G_{11}-G_{12} & 0 \\
0 & 0 & 0 & 0 & 0 & G_{11}-G_{12} 
\end{array}
\right].
\label{eq:kubGiso}
\end{equation}
For interpretation of further calculations, it is convenient to introduce new terms $\Delta G$ and $\Delta \Gamma$
\begin{align}
\Delta G &= G_{11}-2G_{44}-\frac{G_{12}+G_{21}}{2} 
\\
\Delta\Gamma&=\frac{G_{12}-G_{21}}{2}\,.  
\end{align}
Using those terms, the components $G_{12}$ and $G_{21}$ are then expressed as
$G_{12} = G_{11}-2G_{44}-\Delta G+\Delta\Gamma$ and
$G_{21} = G_{11}-2G_{44}-\Delta G-\Delta\Gamma$.  
In cubic crystals with higher symmetry (the crystal classes $\overline{4}3m$, 432
and $m3m$) the parameter $\Delta \Gamma$ vanishes, $\Delta \Gamma=0$. Furthermore, in case of isotropic medium also the parameter $\Delta G$ vanishes, $\Delta G=0$.

\section{Transformation of vectors and second rank tensors under orthogonal transformation of coordinate system}

Now we want to describe the variation of magneto-optical permittivity tensor on the rotation of coordinate system. The original coordinate system (later called ``crystal coordinate system'') defines $xyz$ axes being parallel to the principal axes of crystal (Fig.~\ref{f:one}). 

\begin{figure}[t]
\centering
\includegraphics[scale=0.57,angle=0]{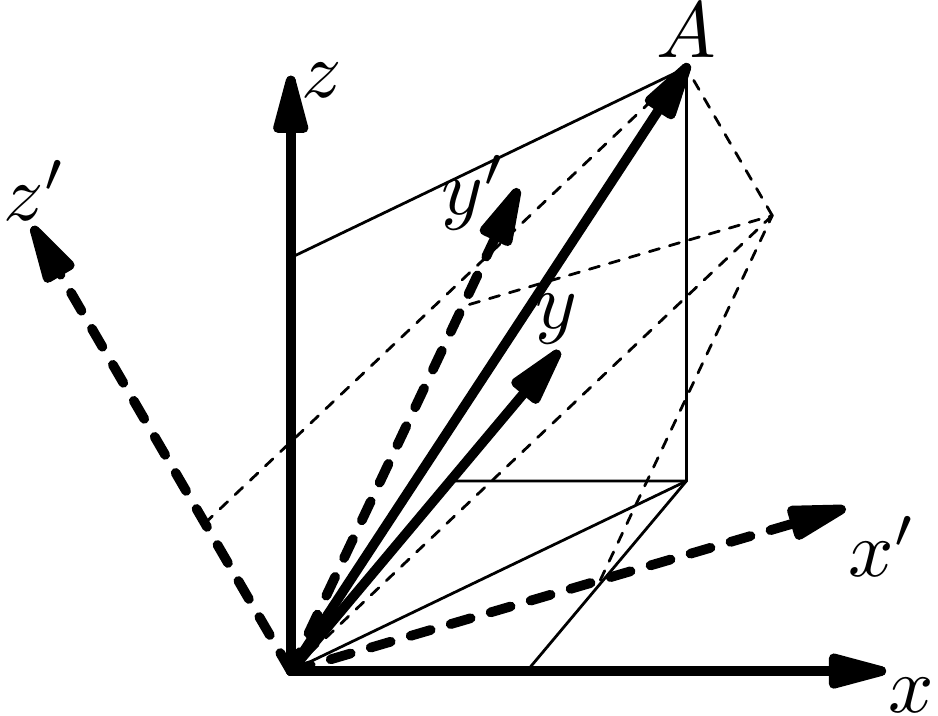}
\caption{\label{f:two}%
Transformation of vector's components under the orthogonal transformation of coordinate system.}
\end{figure}

At first let's suppose that we transform crystal coordinate system according to an orthogonal transformation given by matrix 
${\mathbf R}$ (see Fig.~\ref{f:two}).
Having a vector ${\mathbf A}$ in crystal coordinates $xyz$, then the same vector in new coordinates $x'y'z'$ is expressed by components $\mathbf{A}'$ \cite{birss} (in Einstein notation)
\begin{equation}
A'_{i}=R_{ij}A_j
\label{eq:ARA}
\end{equation}
or written equivalently as ${\mathbf A}'={\mathbf R}\cdot{\mathbf A}$ in matrix notation.
Next, we transform the components of the second rank tensor $\varepsilon_{ij}$ \cite{birss}
 \begin{equation}
\varepsilon'_{ij}=R_{ik}R_{jl}\varepsilon_{kl}\,.
\label{ttransform}
\end{equation}
or equivalently written as $\mathbf{\varepsilon}'=\mathbf{R}\cdot\mathbf{\varepsilon}\cdot\mathbf{R}^{-1}$ in matrix notation.

To perform two orthogonal transformations in succession, we have first to transform coordinate system according to
the matrix ${\mathbf R}^{(1)}$ and then according to the matrix ${\mathbf R}^{(2)}$.
Then the coordinates of the vector ${\mathbf A}$ in new coordinate system are given by complete transformation $R_{ij}=R_{ik}^{(2)}R_{kj}^{(1)}$ (i.e.\ in matrix form $\mathbf{R}=\mathbf{R}^{(2)}\mathbf{R}^{(1)}$).

\section{Crystal in a hypothetical experimental setup}

Now, let us place the cubic crystal (i.e.\ the sample) into a hypothetical setup, described by cartesian axes $x''y''z''$, later called ``sample coordinate system''. Then, the parameters related with the setup, such as e.g.\ the external field direction, plane of incidence of light, flow of charge current, sample surface orientation etc.\ are described within this setup cartesian system. Furthermore, without loss of generality, let us attribute $z''$-axis to be normal to the sample surface and hence $x''$ and $y''$ directions are within sample plane. 

As wrote above, the main crystal axis (100), (010) and (001) are related with $xyz$ coordinate system called ``crystal coordinate system'', in which the quadratic response is described by Eqs.~(\ref{eq:kubG}--\ref{eq:kubGiso}).

\begin{figure}
\centering
\includegraphics[scale=0.5,angle=0]{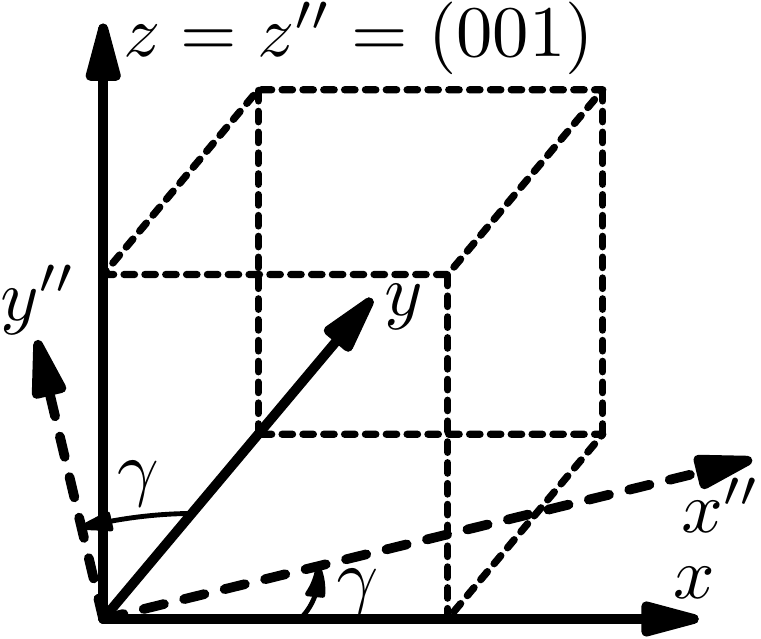}
\caption{\label{f:three}%
Transformation (i.e.\ rotation by angle $\gamma$) of coordinate system for sample with (001) oriented surface. Presented angle $\gamma$ is positive.}
\end{figure}

In following Sections, we describe quadratic response of cubic crystal, where sample surface (i.e.\ sample normal) has orientations (001), (011) or (111). Hence, as the sample surface orientation is fixed within crystal coordinate system, the only free parameter is rotation of the sample around the sample normal, i.e.\ rotation around $z''$ axis by an angle $\gamma$. 
In real experimental investigations, it corresponds to rotation of the sample (in case of optical investigations), or to change of probing in-plane current direction (in case of transport investigations). The positive $\gamma$ direction is defined as follow: when looking into the positive $z''$ direction, the rotation is counter-clock-wise (Fig.~\ref{f:three}).

Finally, recall that the relation between both coordinate systems is given by matrix $\mathbf{R}$ as in Eq.~(\ref{eq:ARA}).

\begin{table*}
\begin{tabular}{|c||c|c|c|c|}
\hline
 \textbf{(001)} 
 & $\ds \ep^{(001)}_{yz}= \ep^{(001)}_{zy}$ 
 & $\ds\ep^{(001)}_{xz}=\ep^{(001)}_{zx}$ 
 & $\ds\ep^{(001)}_{xy}=\ep^{(001)}_{yx}$ 
 & $\ds\ep^{(001)}_{xx}-\ep^{(001)}_{yy}$
 \\ \hline
 $\ds M''_yM''_z$
 & $\ds 2G_{44}$
 & $\ds 0$
 & $\ds 0$
 & $\ds 0$
 \\ \hline
 $\ds M''_xM''_z$
 & $\ds 0$
 & $\ds 2G_{44}$ 
 & $\ds 0$
 & $\ds 0$
 \\ \hline
 $\ds M''_xM''_y$
 & $\ds 0$
 & $\ds 0$
 & $\ds 2 G_{44} +\frac{1}{2}\Delta G (1-\cos4\gamma) $
 & $\ds -\Delta G \sin 4\gamma$
 \\ \hline
 $\ds {M''_y}^2-{M''_z}^2$
 & $\ds 0$
 & $\ds 0$
 & $\ds -\frac{1}{2}\Delta\Gamma \sin2\gamma$
 & $\ds \Delta\Gamma \cos2\gamma$
 \\ \hline
 $\ds {M''_z}^2-{M''_x}^2$
 & $\ds 0$
 & $\ds 0$
 & $\ds \frac{1}{2}\Delta\Gamma \sin2\gamma$
 & $\ds -\Delta\Gamma \cos2\gamma$
 \\ \hline
 $\ds {M''_x}^2-{M''_y}^2$
 &$\ds 0$
 & $\ds 0$
 & $\ds -\frac{1}{4}\Delta G \sin 4\gamma $
 &$\ds 2G_{44}+\Delta G \cos^{2}2\gamma$
 \\ \hline
 \end{tabular}
 \caption{\label{t:001MMoff}%
Off-diagonal elements and AMR-like parameter of the second-order magneto-optical permittivity tensor for cubic crystal with (001) oriented surface.}
\end{table*}
 
\begin{table*}
\begin{tabular}{|c||>{\centering}m{0.283\textwidth}|>{\centering}m{0.283\textwidth}|>{\centering}m{0.25\textwidth}|}
 \hline
  \textbf{(001)} 
  & $\ds\ep^{(001)}_{xx}$ & $\ds\ep^{(001)}_{yy}$ & $\ds\ep^{(001)}_{zz}$
 \tabularnewline\hline
 $\ds {M''_y} {M''_z}$ & 0 &0 & 0
 \tabularnewline \hline
 $\ds {M''_x} {M''_z}$ & 0 & 0& 0
 \tabularnewline \hline
 $\ds {M''_x} {M''_y}$ 
 &  $\ds -\frac{\Delta G}{2}\sin4\gamma + \Delta\Gamma\sin2\gamma$
 &  $\ds \frac{\Delta G}{2}\sin4\gamma + \Delta\Gamma\sin2\gamma$
 & $\ds -2\Delta\Gamma \sin2\gamma$
 \tabularnewline \hline
 $\ds {M''_x}^2$ 
 & $\ds G_{11} \ds -\frac{1}{2}\Delta G \sin^{2}2\gamma$ 
 & $\ds  G_{11}-2G_{44}-\frac{1}{2}\Delta G(1+\cos^{2}2\gamma)
   \newline -\Delta\Gamma\cos2\gamma$ 
 & $ \ds G_{11}-2G_{44}-\Delta G +\Delta\Gamma\cos2\gamma$
 \tabularnewline \hline
 $\ds {M''_y}^2$ 
 & $\ds G_{11}-2G_{44}-\frac{1}{2}\Delta G(1+\cos^{2}2\gamma) 
 \newline
 +\Delta\Gamma\cos2\gamma $
& $\ds G_{11} \ds -\frac{1}{2}\Delta G \sin^{2}2\gamma$ 
  & $ \ds G_{11}-2G_{44}-\Delta G -\Delta\Gamma\cos2\gamma$
 \tabularnewline \hline
 $\ds {M''_z}^2$ 
 &  $ \ds G_{11}-2G_{44}-\Delta G -\Delta\Gamma\cos2\gamma$
 & $ \ds G_{11}-2G_{44}-\Delta G +\Delta\Gamma\cos2\gamma$
 & $\ds G_{11}$
 \tabularnewline \hline
 \end{tabular}
 \caption{\label{t:001MMdiag}%
Diagonal elements of second-order magneto-optical permittivity tensor for cubic crystal with (001) oriented surface.}
\end{table*}

 \begin{table*}
 \begin{tabular}{|>{\centering}m{0.12\textwidth}||m{0.65\textwidth}|}
 \hline
 $\ds \ep^{(001)}_{yz}=\ep^{(001)}_{zy}$ & $\ds G_{44} \sin 2\theta \sin\varphi$
 \\ \hline
 $\ds \ep^{(001)}_{xz}=\ep^{(001)}_{zx}$ & $\ds G_{44}\sin2\theta \cos\varphi$
 \\ \hline
 $\ds \ep^{(001)}_{xy}=\ep^{(001)}_{yx}$ & $\ds G_{44} \sin2\varphi \sin^{2}\theta - \frac{1}{2}\Delta G \sin2\gamma \cos(2\gamma+2\varphi)\sin^{2}\theta
 + \frac{1}{4}\Delta \Gamma \sin2\gamma (3\cos2\theta+1)$
 \\ \hline
 $\ds \ep^{(001)}_{xx}$ &
 $\ds G_{11}-2G_{44}(1-\cos^{2}\varphi \sin^{2}\theta) 
 + \frac{1}{4}\Delta G [\sin^{2}\theta\cos(4\gamma+2\varphi)+\sin^{2}\theta\cos2\varphi+
2\sin^{2}\theta-4]
\newline
-\frac{1}{2}\Delta\Gamma[\sin^{2}\theta\cos(2\gamma+2\varphi)+\cos2\gamma(2-3\sin^2\theta)]$
 \\ \hline
 $\ds\ep^{(001)}_{yy}$ &
 $\ds G_{11}-2G_{44}(1-\sin^{2}\varphi\sin^{2}\theta)-\frac{1}{4}\Delta G[\sin^{2}\theta\cos(4\gamma+2\varphi)
+\cos2\varphi\sin^{2}\theta-2\sin^{2}\theta+4]
\newline
-\frac{1}{2}\Delta\Gamma[\sin^{2}\theta\cos(2\gamma+2\varphi)+\cos2\gamma(3\sin^{2}\theta-2)]$
 \\ \hline
 $\ds\ep^{(001)}_{zz}$ &
 $\ds G_{11}-2G_{44}\sin^{2}\theta-\Delta G\sin^{2}\theta+\Delta\Gamma\sin^{2}\theta\cos(2\gamma+2\varphi)$
 \\ \hline
 $\ds \ep^{(001)}_{xx}-\ep^{(001)}_{yy}$ &
 $\ds  2G_{44}\sin^2\theta\cos2\varphi 
 + \Delta G\sin^2\theta\cos2\gamma\cos(2\gamma+2\varphi)
 + \Delta\Gamma\cos2\gamma(1-3\cos^2\theta)$
 \\ \hline
 \end{tabular}
 \caption{\label{t:001Mpolar}%
Elements of the second-order magneto-optical permittivity tensor for cubic crystal with (001) oriented surface, with magnetization vector expressed in polar coordinates $\varphi$ and $\theta$.}
\end{table*}

\begin{table*}
\addtolength{\tabcolsep}{-3pt}
\begin{tabular}{|c||c|c|c|c|}
\hline
 \textbf{(011)} 
 & $\ds \ep^{(011)}_{yz}= \ep^{(011)}_{zy}$ 
 & $\ds\ep^{(011)}_{xz}=\ep^{(011)}_{zx}$ 
 & $\ds\ep^{(011)}_{xy}=\ep^{(011)}_{yx}$ 
 & $\ds\ep^{(011)}_{xx}-\ep^{(011)}_{yy}$
 \\ \hline
   $\ds M''_yM''_z$
 & $\ds 2G_{44}+\frac{1}{2}\Delta G(1+\cos2\gamma)$
 & $\ds \frac{1}{2}\Delta G\sin2\gamma$
 & $\ds -\frac{3}{2}\Delta \Gamma\sin\gamma(1+\cos2\gamma)$
 & $\ds 3\Delta \Gamma\cos\gamma\cos2\gamma$
 \\ \hline
    $\ds M''_xM''_z$
 &  $\ds \frac{1}{2}\Delta G\sin2\gamma$
 &  $\ds 2G_{44}+\frac{1}{2}\Delta G(1-\cos2\gamma)$
 &  $\ds -\frac{3}{2}\Delta \Gamma\cos\gamma(1-\cos2\gamma)$
 &  $\ds 3\Delta \Gamma\sin\gamma\cos2\gamma$
 \\ \hline
    $\ds M''_xM''_y$
 &  $\ds \frac{3}{2}\Delta \Gamma\sin\gamma(1+\cos2\gamma)$
 &  $\ds \frac{3}{2}\Delta \Gamma\cos\gamma(1-\cos2\gamma)$
 &  $\ds 2G_{44}+\frac{3}{8}\Delta G(1-\cos4\gamma)$
 &  $\ds -\frac{3}{4}\Delta G\sin4\gamma$
 \\ \hline
   $\ds {M''_y}^2-{M''_z}^2$
 & $\ds -\Delta \Gamma\cos\gamma$
 & $\ds \frac{1}{2}\Delta \Gamma\sin\gamma$
 & $\ds \frac{1}{4}\Delta G\sin2\gamma$
 & $\ds \frac{1}{4}\Delta G\cos2\gamma$
 \\ \hline
   $\ds {M''_z}^2-{M''_x}^2$
 & $\ds -\frac{1}{2}\Delta \Gamma\cos\gamma$
 & $\ds \Delta \Gamma\sin\gamma$
 & $\ds \frac{1}{2}\Delta G\sin2\gamma$
 & $\ds -\frac{1}{4}\Delta G\cos2\gamma$
 \\ \hline
    $\ds {M''_x}^2-{M''_y}^2$
 &  $\ds -\frac{3}{2}\Delta \Gamma\cos^3\gamma$
 &  $\ds \frac{3}{2}\Delta \Gamma\sin^3\gamma$
 &  $\ds \frac{3}{4}\Delta G\sin^2\gamma\sin2\gamma$
 &  $\ds 2G_{44}+\frac{3}{8}\Delta G(1+\cos4\gamma)$
 \\ \hline
 \end{tabular}
 \caption{\label{t:011MMoff}%
Off-diagonal elements and AMR-like parameter of second-order magneto-optical permittivity tensor for cubic crystal with (011) oriented surface.}
\end{table*}
 
\begin{table*}
\addtolength{\tabcolsep}{-2.3pt}
 \begin{tabular}{|c||c|c|c|}
 \hline
  \textbf{(011)} 
  & $\ds\ep^{(011)}_{xx}$ & $\ds\ep^{(011)}_{yy}$ & $\ds\ep^{(011)}_{zz}$
 \\ \hline
 $\ds M''_y M''_z$ & $\ds \frac{1}{2}\Delta \Gamma\cos\gamma(1+3\cos2\gamma)$ 
 &  $\ds \frac{1}{2}\Delta \Gamma\cos\gamma(1-3\cos2\gamma)$ &
 $\ds -\Delta \Gamma\cos\gamma$ 
 \\ \hline
 $\ds M''_x M''_z$ & 
 $\ds \frac{1}{2}\Delta \Gamma\sin\gamma(1+3\cos2\gamma)$ & 
 $\ds \frac{1}{2}\Delta \Gamma\sin\gamma(1-3\cos2\gamma)$ & 
 $\ds -\Delta \Gamma\sin\gamma$
 \\ \hline
 $\ds M''_x M''_y$ 
 & $\ds -\frac{1}{8}\Delta G(2\sin2\gamma +3\sin4\gamma)$
 & $\ds -\frac{1}{8}\Delta G(2\sin2\gamma -3\sin4\gamma)$
 & $\ds \frac{1}{2}\Delta G\sin2\gamma$
 \\ \hline
 $\ds {M''_x}^2$ 
 & $\ds G_{11}  - \frac{1}{16}\Delta G(7-4\cos2\gamma-3\cos4\gamma)$ 
 &  $\ds G_{11}  - 2G_{44}-\frac{1}{16}\Delta G(13+3\cos4\gamma)$  
 &  $\ds G_{11}  - 2G_{44}-\frac{1}{4}\Delta G(3+\cos2\gamma)$ 
 \\ \hline
 $\ds {M''_y}^2$ 
 &  $\ds G_{11}  - 2G_{44}-\frac{1}{16}\Delta G(13+3\cos4\gamma)$  
&  $\ds G_{11}  - \frac{1}{16}\Delta G(7+4\cos2\gamma-3\cos4\gamma)$  
  & $\ds  G_{11}  - 2G_{44}-\frac{1}{4}\Delta G(3-\cos2\gamma)$
 \\ \hline
 $\ds {M''_z}^2$ 
 & $ \ds  G_{11}  - 2G_{44}-\frac{1}{4}\Delta G(3+\cos2\gamma)$
 & $\ds G_{11}  - 2G_{44}-\frac{1}{4}\Delta G(3-\cos2\gamma)$ 
 & $\ds  G_{11}-\frac{1}{2}\Delta G$
 \\ \hline
 \end{tabular}
 \caption{\label{t:011MMdiag}%
Diagonal elements of second-order magneto-optical permittivity tensor for cubic crystal with (011) oriented surface.}
\end{table*} 
\begin{table*}
 \begin{tabular}{|>{\centering}m{0.115\textwidth}||>{\centering}m{0.825\textwidth}|}
 \hline
 $\ds \ep^{(011)}_{yz}=\ds \ep^{(011)}_{zy}$ 
 & $\ds G_{44}\sin\varphi\sin2\theta+\frac{1}{2}\Delta G\sin(\gamma+\varphi)\cos\gamma\sin2\theta
+\frac{1}{8}\Delta \Gamma\cos\gamma[3\cos(2\gamma+2\varphi)(\cos2\theta-1)+3\cos2\theta+1]$
 \tabularnewline\hline
 $\ds \ep^{(011)}_{xz}=\ds \ep^{(011)}_{zx}$ 
 &  $\ds G_{44}\cos\varphi\sin2\theta+\frac{1}{2}\Delta G\sin(\gamma+\varphi)\sin\gamma\sin2\theta
+\frac{1}{8}\Delta \Gamma\sin\gamma[3\cos(2\gamma+2\varphi)(\cos2\theta-1)+3\cos2\theta+1]$
 \tabularnewline\hline
 $\ds \ep^{(011)}_{xy}=\ds \ep^{(011)}_{yx}$ 
 & $\ds G_{44}\sin2\varphi\sin^2\theta
 +\frac{1}{16}\Delta G\sin2\gamma[3\cos(2\gamma+2\varphi)(\cos2\theta -1) +3\cos2\theta+1]
-\frac{3}{4}\Delta \Gamma\sin2\gamma\sin(\gamma+\varphi)\sin2\theta
 $
  \tabularnewline\hline
 $\ds\ep^{(011)}_{xx}$ 
 & $\ds G_{11}-2G_{44}(1-\cos^2\varphi\sin^2\theta)
+\frac{1}{4}\Delta\Gamma\sin(\gamma+\varphi)(1+3\cos2\gamma)\sin2\theta
\newline
 +\frac{1}{16}\Delta G[
\cos(2\gamma+2\varphi)(1+3\cos2\gamma)(1-\cos2\theta)-\cos2\gamma(1+3\cos2\theta)-\cos2\theta-11]
$
 \tabularnewline\hline
 $\ep^{(011)}_{yy}$ 
 & $\ds G_{11}-2G_{44}(1-\sin^2\varphi\sin^2\theta)
+\frac{1}{4}\Delta\Gamma\sin(\gamma+\varphi)(1-3\cos2\gamma)\sin2\theta
\newline
 +\frac{1}{16}\Delta G[
\cos(2\gamma+2\varphi)(1-3\cos2\gamma)(1-\cos2\theta)+\cos2\gamma(1+3\cos2\theta)-\cos2\theta-11]
$
 \tabularnewline\hline
 $\ep^{(011)}_{zz}$ 
 & $\ds G_{11}-2G_{44}\sin^2\theta+\frac{1}{8}\Delta G[
\cos(2\gamma+2\varphi)(\cos2\theta-1)+\cos2\theta-5]
-\frac{1}{2}\Delta\Gamma\sin(\gamma+\varphi)\sin2\theta$
 \tabularnewline\hline
 $\ep^{(011)}_{xx}-\ep^{(011)}_{yy}$ 
 & $\ds 2G_{44}\sin^2\theta\cos2\varphi
  + \frac{1}{8}\Delta G \cos2\gamma[3\cos(2\gamma+2\varphi)(1-\cos2\theta)-3\cos2\theta-1]
  +\frac{3}{2}\Delta\Gamma \sin(\gamma+\varphi)\sin2\theta\cos2\gamma$ 
 \tabularnewline\hline
 \end{tabular}
 \caption{\label{t:011Mpolar}%
Elements of second-order magneto-optical permittivity tensor for cubic crystal with (011) oriented surface, with magnetization vector expressed in polar coordinates $\varphi$ and $\theta$.}
\end{table*}
\begin{table*}
 \begin{tabular}{@{\extracolsep{-4pt}}|>{\centering}m{0.093\textwidth}||>{\centering}m{0.201\textwidth}|>{\centering}m{0.201\textwidth}|>{\centering}m{0.201\textwidth}|>{\centering}m{0.201\textwidth}|}
\hline
  \textbf{(111)} 
 & $\ds \ep^{(111)}_{yz}= \ep^{(111)}_{zy}$ 
 & $\ds\ep^{(111)}_{xz}=\ep^{(111)}_{zx}$ 
  & $\ds\ep^{(111)}_{xy}=\ep^{(111)}_{yx}$ 
 & $\ds\ep^{(111)}_{xx}-\ep^{(111)}_{yy}$
 \tabularnewline\hline
 $\ds {M''_y}{M''_z}$
 & $\ds 2G_{44}+\frac{2}{3}\Delta G$
 &  $\ds\frac{2\sqrt{3}}{3}\Delta \Gamma$
  & $\ds\frac{1}{3}\Delta G(\sin3\gamma-\cos3\gamma)
-\frac{\sqrt{3}}{3}\Delta\Gamma(\sin3\gamma+\cos3\gamma) $
 &  $\ds-\frac{2}{3}\Delta G(\cos3\gamma+\sin3\gamma)
+\frac{2\sqrt{3}}{3}\Delta\Gamma(\cos3\gamma-\sin3\gamma) $
 \tabularnewline\hline
 $\ds {M''_x}{M''_z}$
 & $\ds-\frac{2\sqrt{3}}{3}\Delta \Gamma$
 &  $\ds2G_{44}+\frac{2}{3}\Delta G$
  &  $\ds-\frac{1}{3}\Delta G(\sin3\gamma+\cos3\gamma)
+\frac{\sqrt{3}}{3}\Delta\Gamma(\cos3\gamma-\sin3\gamma) $
 &   $\ds\frac{2}{3}\Delta G(\cos3\gamma-\sin3\gamma)
+\frac{2\sqrt{3}}{3}\Delta\Gamma(\cos3\gamma+\sin3\gamma) $
 \tabularnewline\hline
 $\ds {M''_x}{M''_y}$
 &  $\ds\frac{1}{3}\Delta G(\sin3\gamma-\cos3\gamma)
+\frac{\sqrt{3}}{3}\Delta\Gamma(\sin3\gamma+\cos3\gamma) $
 &   $\ds-\frac{1}{3}\Delta G(\sin3\gamma+\cos3\gamma)
+\frac{\sqrt{3}}{3}\Delta\Gamma(\sin3\gamma-\cos3\gamma) $
 &  $\ds2G_{44}+\frac{1}{3}\Delta G$
 &  $\ds-\frac{2\sqrt{3}}{3}\Delta\Gamma$
 \tabularnewline\hline
 $\ds {M''_y}^2-{M''_z}^2$
 & $0$
 & $0$
  & $0$
 &  $0$
 \tabularnewline\hline
 $\ds {M''_z}^2-{M''_x}^2$
 & $0 $
 & $0$
  & $0$
 &  $0$
 \tabularnewline\hline
 $\ds {M''_x}^2-{M''_y}^2$
 & $\ds-\frac{1}{6}\Delta G(\sin3\gamma+\cos3\gamma)
+\frac{\sqrt{3}}{6}\Delta\Gamma(\sin3\gamma-\cos3\gamma) $
 &  $\ds\frac{1}{6}\Delta G(\cos3\gamma-\sin3\gamma)
-\frac{\sqrt{3}}{6}\Delta\Gamma(\sin3\gamma+\cos3\gamma) $
 & $\ds\frac{\sqrt{3}}{6}\Delta\Gamma$
 &  $\ds2G_{44}+\frac{1}{3}\Delta G$
 \tabularnewline\hline
\end{tabular}
\caption{\label{t:111MMoff}%
Off-diagonal elements and AMR-like parameter of second-order magneto-optical permittivity tensor for cubic crystal with (111) surface orientation.
}
\end{table*}

\section{Quadratic magneto-optical permittivity tensor for cubic crystal with (001) oriented surface}

Let's suppose a cubic crystal grown with (001) surface direction. Hence, the setup coordinate system and the crystal coordinate system have equal $z$-axes, $z=z''$ pointing in sample surface normal direction. Then, the rotation around angle $\gamma$ is described by transformation matrix ${\mathbf R}_z(\gamma)$ (Fig.~\ref{f:three})
\begin{equation}
{\mathbf R}^{(001)}={\mathbf R}_z(\gamma)=\left[ 
\begin{array}{ccc}
\cos\gamma & \sin\gamma & 0 \\
-\sin\gamma & \cos\gamma & 0 \\
0 & 0 & 1
\end{array}
\right]\,.
\label{Rz}
\end{equation}  
Before expressing the elements of permittivity tensor $\varepsilon_{ij}^{(2)}$ given by (\ref{eq:kubG}--\ref{eq:kubGiso}) in the setup coordinate system $x''y''z''$, $\varepsilon_{ij}^{(001)}$, first we need to express components of magnetization vector in the crystal coordinate system. 
It originates from the fact that during hypothetical measurements, we describe magnetization vector ${\mathbf M}''$ in setup coordinate system $x''y''z''$, but elements $M_k$, $M_l$ coming in Eqs.~(\ref{eq:kubG}--\ref{eq:kubGiso}) are components of the same magnetization vector $\mathbf{M}$ expressed in the crystal coordinate system $xyz$.
Recall, the relation of the magnetization vector between both coordinate system is (Eq.~(\ref{eq:ARA}))
\begin{equation}
M_{k}={R^{(001)}}^{-1}_{ks}M''_s = R_{sk}^{(001)} M''_s\,.
\end{equation}
Hence the permittivity tensor in crystal coordinate system $xyz$ with magnetizations from setup coordinate system $x''y''z''$ is
\begin{equation} 
\varepsilon_{ij}^{(2)}=G_{ijkl}{R^{(001)}}^{-1}_{ks}M''_s{R^{(001)}}^{-1}_{lt}M''_t\,.
\label{epold}
  \end{equation}
Finally, transformation of the permittivity elements $\varepsilon_{ij}^{(2)}$ into the sample coordinate system    is (Eq.~\ref{ttransform})
\begin{equation}
\varepsilon^{(001)}_{ij}=R^{(001)}_{ik}R^{(001)}_{jl}\varepsilon^{(2)}_{kl}\,.
\label{epnewtmp}
\end{equation}
which leads to the final expression of the quadratic magneto-optical tensor in the setup coordinate system, $\varepsilon^{(001)}_{ij}$, with magnetization direction also described in the setup coordinate system
\begin{equation}
\varepsilon^{(001)}_{ij}=R^{(001)}_{ik}R^{(001)}_{jl}
G_{klmn}{R^{(001)}}^{-1}_{ms}M''_s{R^{(001)}}^{-1}_{nt}M''_t
\,.
\label{epnew}
\end{equation}
where recall $G_{klmn}$ are elements of the quadratic magneto-optical tensor, as described by Eqs.(\ref{eq:kubG}--\ref{eq:kubGiso}). 
In Tabs.~\ref{t:001MMoff} and~\ref{t:001MMdiag} are expressed calculated off-diagonal and diagonal elements of $\varepsilon^{(001)}_{ij}$, as well as AMR-like parameter $\ep^{(001)}_{xx}-\ep^{(001)}_{yy}$. In corresponding rows are separated terms proportional to magnetization directions
 $M''_yM''_z$,  $M''_xM''_z$,  $M''_xM''_y$ and  $ {M''_y}^2-{M''_z}^2$,  $ {M''_z}^2-{M''_x}^2$,  ${M''_x}^2-{M''_y}^2$.

For comparison of experimental or ab-initio results with symmetry arguments, it is often useful to express magnetization vector ${\mathbf M}''$ in spherical coordinates
\begin{equation}
\mathbf{M}''=\begin{bmatrix}
M''_x \\ M''_y \\ M''_z
\end{bmatrix}
=
\begin{bmatrix}
\sin\theta \cos\varphi \\
\sin\theta \sin\varphi \\
\cos\theta
\end{bmatrix}\,.
\label{eq:Mdef}
\end{equation}
Then, all permittivity elements $\varepsilon^{(001)}_{ij}$ can be expressed to depend on angles $\varphi$ and $\theta$ as presented in Tab.~\ref{t:001Mpolar}.
\clearpage

\begin{table*}
 \begin{tabular}{|>{\centering}m{0.07\textwidth}||>{\centering}m{0.22\textwidth}|>{\centering}m{0.22\textwidth}|>{\centering}m{0.22\textwidth}|}
 \hline
   \textbf{(111)} 
  & $\ds\ep^{(111)}_{xx}$ & $\ds\ep^{(111)}_{yy}$ & $\ds\ep^{(111)}_{zz}$
 \tabularnewline\hline
 $\ds{M''_y} {M''_z}$ &
 $\ds-\frac{1}{3}\Delta G(\cos3\gamma+\sin3\gamma)
+\frac{\sqrt{3}}{3}\Delta\Gamma(\cos3\gamma-\sin3\gamma) $ &
  $\ds\frac{1}{3}\Delta G(\cos3\gamma+\sin3\gamma)
+\frac{\sqrt{3}}{3}\Delta\Gamma(\sin3\gamma-\cos3\gamma) $ & 
$  0 $ 
 \tabularnewline\hline
 $\ds{M''_x} {M''_z}$ & 
 $\ds\frac{1}{3}\Delta G(\cos3\gamma-\sin3\gamma)
+\frac{\sqrt{3}}{3}\Delta\Gamma(\cos3\gamma+\sin3\gamma) $ & 
 $\ds\frac{1}{3}\Delta G(\sin3\gamma-\cos3\gamma)
-\frac{\sqrt{3}}{3}\Delta\Gamma(\cos3\gamma+\sin3\gamma) $ & 
$  0 $
 \tabularnewline\hline
 $\ds{M''_x} {M''_y}$ 
 &   $\ds-\frac{\sqrt{3}}{3}\Delta\Gamma $ 
 &  $\ds\frac{\sqrt{3}}{3}\Delta\Gamma $ 
 &  $ 0 $ 
 \tabularnewline\hline
 $\ds {M''_x}^2$ 
 & $\ds G_{11}-\frac{1}{2}\Delta G$ 
 &  $\ds  G_{11}-2G_{44}-\frac{5}{6}\Delta G $
 &  $\ds  G_{11}-2G_{44}-\frac{2}{3}\Delta G$
 \tabularnewline\hline
 $\ds{M''_y}^2$ 
 & $\ds  G_{11}-2G_{44}-\frac{5}{6}\Delta G $ 
&  $\ds G_{11}-\frac{1}{2}\Delta G$ 
  &  $\ds  G_{11}-2G_{44}-\frac{2}{3}\Delta G$
 \tabularnewline\hline
 $\ds{M''_z}^2$ 
 &  $\ds  G_{11}-2G_{44}-\frac{2}{3}\Delta G$ 
 &  $\ds  G_{11}-2G_{44}-\frac{2}{3}\Delta G$
 &  $\ds  G_{11}-\frac{2}{3}\Delta G$
 \tabularnewline\hline
 \end{tabular}
\caption{\label{t:111MMdiag}%
Diagonal elements of second-order magneto-optical permittivity tensor for cubic crystal with (111) surface orientation.}
\end{table*}

\begin{table*}
 \begin{tabular}{|>{\centering}m{0.12\textwidth}||>{\centering}m{0.82\textwidth}|}
 \hline
 $\ds\ep^{(111)}_{yz}$ 
 & $\ds G_{44}\sin\varphi\sin2\theta
 +\frac{1}{12}\Delta G \{[\cos(3\gamma+2\varphi)+\sin(3\gamma+2\varphi)](\cos2\theta-1)
+4\sin\varphi\sin2\theta\}
\newline
+\frac{\sqrt{3}}{12}\Delta\Gamma\{[\cos(3\gamma+2\varphi)-\sin(3\gamma+2\varphi)](\cos2\theta-1)-4\cos\varphi\sin2\theta\}
$
 \tabularnewline\hline
 $\ds\ep^{(111)}_{xz}$ 
 & $\ds G_{44}\cos\varphi\sin2\theta
+\frac{1}{12}\Delta G \{[\cos(3\gamma+2\varphi)-\sin(3\gamma+2\varphi)](1-\cos2\theta)
 +4\cos\varphi\sin2\theta\}
\newline
 +\frac{\sqrt{3}}{12}\Delta\Gamma\{
[\cos(3\gamma+2\varphi)+\sin(3\gamma+2\varphi)](\cos2\theta-1)+4\sin\varphi\sin2\theta\}
$
 \tabularnewline\hline
 $\ds \ep^{(111)}_{xy}$ 
 & $\ds G_{44}\sin2\varphi\sin^2\theta
 -\frac{1}{12}\Delta G \{2[\cos(3\gamma+\varphi)+\sin(3\gamma+\varphi)]\sin2\theta
+\sin2\varphi(\cos2\theta-1)\}
\newline
+\frac{\sqrt{3}}{12}\Delta\Gamma\{
2[\cos(3\gamma+\varphi)-\sin(3\gamma+\varphi)]\sin2\theta+\cos2\varphi(1-\cos2\theta)\}
$
 \tabularnewline\hline
 $\ds\ep^{(111)}_{xx}$ 
 & $\ds G_{11} -2G_{44}(1-\cos^2\varphi\sin^2\theta)
 +\frac{1}{12}\Delta G \{\cos2\varphi(1-\cos2\theta)
+2[\cos(3\gamma+\varphi)-\sin(3\gamma+\varphi)]\sin2\theta-8\}
\newline
 +\frac{\sqrt{3}}{12}\Delta \Gamma \{\sin2\varphi(\cos2\theta-1)
+2[\cos(3\gamma+\varphi)+\sin(3\gamma+\varphi)]\sin2\theta\}
$
 \tabularnewline\hline
 $\ds \ep^{(111)}_{yy}$ 
 & $\ds G_{11}-2G_{44}(1-\sin^2\varphi\sin^2\theta)
 +\frac{1}{12}\Delta G \{\cos2\varphi(\cos2\theta-1)
+2[\sin(3\gamma+\varphi)-\cos(3\gamma+\varphi)]\sin2\theta-8\}
\newline
 +\frac{\sqrt{3}}{12}\Delta \Gamma \{\sin2\varphi(1-\cos2\theta)
-2[\cos(3\gamma+\varphi)+\sin(3\gamma+\varphi)]\sin2\theta\}
$
 \tabularnewline\hline
 $\ds \ep^{(111)}_{zz}$ 
 & $\ds G_{11}-2G_{44}\sin^2\theta-\frac{2}{3}\Delta G $
 \tabularnewline\hline
$\ds\ep^{(111)}_{xx}-\ep_{yy}^{(111)}$
& $\ds 2G_{44}\sin^2\theta\cos2\varphi
+ \frac{1}{6}\Delta G\left[ 
\cos2\varphi(1-\cos2\theta)+2\sin2\theta[\cos(3\gamma+\varphi)-\sin(3\gamma+\varphi)]
\right]
\newline
+\frac{\sqrt{3}}{6}\Delta\Gamma \left[
(\cos2\theta-1)\sin2\varphi + 2\sin2\theta[\cos(3\gamma+\varphi)+\sin(3\gamma+\varphi)]
\right]
$
 \tabularnewline\hline
 \end{tabular}
  \caption{\label{t:111Mpolar}%
Elements of second-order magneto-optical permittivity tensor for cubic crystal with  surface  perpendicular to the (111) direction 
with magnetization vector expressed in polar coordinates $\varphi$ and $\theta$.}
\end{table*}

\section{Quadratic magneto-optical permittivity tensor for cubic crystal with (011) oriented surface}

\begin{figure}[h]
\centering
\includegraphics[scale=0.5,angle=0]{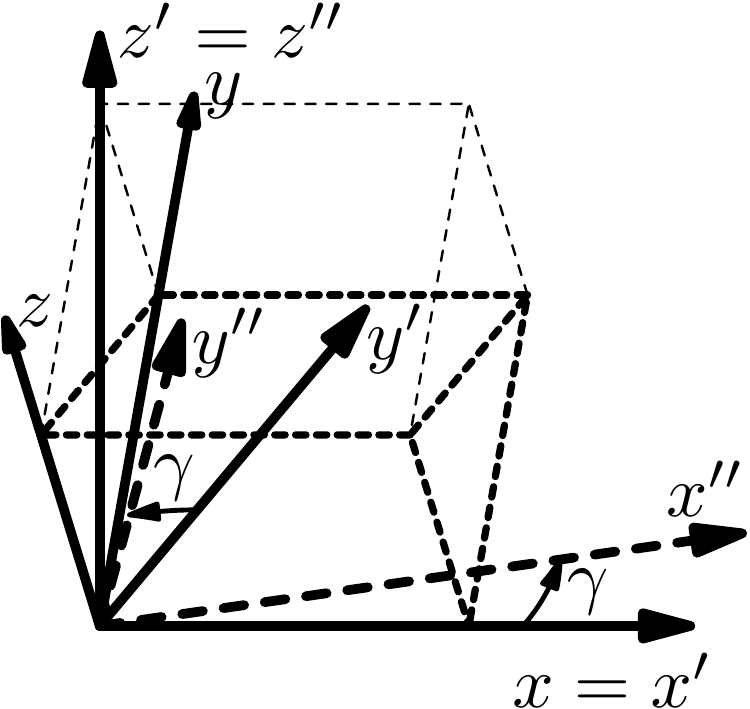}
\caption{\label{f:four}%
Transformation of coordinate system for sample with (011) oriented surface.}
\end{figure}

Let us discuss a cubic crystal grown with (011) surface orientation, i.e.\ the  normal of the sample surface is parallel to the [011] direction. 
First we define a new coordinate system $x'y'z'$, corresponding to cubic crystal oriented (011) (Fig.~\ref{f:four}), where $z'$ is parallel to sample surface normal. 
To obtain new coordinate system $x'y'z'$ we rotate old coordinate system $xyz$ about angle $-\pi/4$ about $x$-axis (Fig.~\ref{f:four}).  
This is described by transformation matrix ${\mathbf R}_x(-\pi/4)$
\begin{equation}
{\mathbf R}_x(-\pi/4)=\left[ 
\begin{array}{ccc}
1 & 0 & 0 \\
0 & \sqrt{2}/2 & -\sqrt{2}/2 \\
0 & \sqrt{2}/2 & \sqrt{2}/2
\end{array}
\right]\,.
\label{rotx}
\end{equation}  
Furthermore, we introduce rotation of the sample by an angle $\gamma$ around the sample normal, i.e.\ rotation of 
$x'y'$ plane around $z'=z''$-axis by the angle $\gamma$, providing setup cartesian system $x''y''z''$. This corresponds to sample rotation by its normal or to change of direction of probing in-plane current. 
This transformation from coordinates $x'y'z'$ to coordinates $x''y''z''$ is given again by
matrix ${\mathbf R}_z(\gamma)$ (Eq.~(\ref{Rz})).
The total transformation matrix ${\mathbf R}^{(011)}$ from $xyz$ crystal coordinates to $x''y''z''$ setup coordinates is 
\begin{equation}
{\mathbf R}^{(011)}={\mathbf R}_z(\gamma){\mathbf R}_x(-\pi/4)\,.
\end{equation}
The quadratic magneto-optical permittivity tensor transforms similarly as in previous Section (Eq.~(\ref{epnew}))
\begin{equation}
\varepsilon^{(011)}_{ij}=R^{(011)}_{ik}R^{(011)}_{jl}  
G_{klmn}{R^{(011)}}^{-1}_{ms}M''_s{R^{(011)}}^{-1}_{nt}M''_t
\,.
\label{ttransform011}
\end{equation} 
The calculated off-diagonal elements of $\varepsilon^{(011)}_{ij}$ and AMR-like parameter $\ep^{(011)}_{xx}-\ep^{(011)}_{yy}$ are presented in Tab.~\ref{t:011MMoff}, as a function of quadratic form of magnetization components $M''_k M''_l$.
The calculated diagonal elements of $\varepsilon^{(011)}_{ij}$ are presented in Tab.~\ref{t:011MMdiag}.  
Finally, Tab.~\ref{t:011Mpolar} provides all elements  of $\varepsilon^{(011)}_{ij}$ expressed as a function of spherical coordinates of magnetization ${\mathbf M}''$, using angles $\varphi$ and $\theta$, as defined in Eq.~(\ref{eq:Mdef}).

\section{Quadratic magneto-optical permittivity tensor for cubic crystal with (111) oriented surface}

Let us discuss the cubic crystal with (111) oriented surface. Then, the $z'$-axis, being perpendicular to the sample surface, is parallel to the [111] direction.
To obtain new coordinate system $x'y'z'$ we rotate crystal coordinate system $xyz$ about angle $\alpha$ around $(-xy)$-axis, where $\cos\alpha=\frac{1}{\sqrt{3}}$ and $\sin\alpha=\frac{\sqrt{2}}{\sqrt{3}}$.  

To calculate transformation matrix we use well-known transformation matrix $\mathbf{R}_{\vec{n}}(\alpha)$ for rotation about arbitrary angle $\alpha$ about arbitrary axis given by normalized vector $\mathbf{n}=(n_1,n_2,n_3)$ 

\begin{widetext}
\begin{equation}
\mathbf{R}_{\vec{n}}(\alpha)=\left[
\begin{array}{ccc}
\cos\alpha+n_1^2(1-\cos\alpha) &  n_1n_2(1-\cos\alpha)+n_3\sin\alpha & n_1n_3(1-\cos\alpha)-n_2\sin\alpha\\
n_1n_2(1-\cos\alpha)-n_3\sin\alpha & \cos\alpha+n_2^2(1-\cos\alpha) & n_2n_3(1-\cos\alpha)+n_1\sin\alpha\\
n_1n_3(1-\cos\alpha)+n_2\sin\alpha & n_2n_3(1-\cos\alpha)-n_1\sin\alpha & \cos\alpha+n_3^2(1-\cos\alpha)
\end{array}
\right]
\end{equation}
\end{widetext}
In our case $(n_1,n_2,n_3)=(-\frac{1}{\sqrt{2}},\frac{1}{\sqrt{2}},0)$ and $\sin\alpha=\frac{1}{\sqrt{3}}$ we have
\begin{equation}
\mathbf{R}_{(-xy)}(\alpha)=\left[
\begin{array}{ccc}
\frac{1}{2}+\frac{1}{2\sqrt{3}} &  -\frac{1}{2}+\frac{1}{2\sqrt{3}} & -\frac{1}{\sqrt{3}}\\
-\frac{1}{2}+\frac{1}{2\sqrt{3}} & \frac{1}{2}+\frac{1}{2\sqrt{3}} &  -\frac{1}{\sqrt{3}}\\
 \frac{1}{\sqrt{3}} &  \frac{1}{\sqrt{3}} & \frac{1}{\sqrt{3}}
\end{array}
\right]\,.
\end{equation}
Next step is to rotate sample by an angle $\gamma$ around $z'=z''$ axis about arbitrary angle $\gamma$ given by transformation matrix
$\mathbf{R}_z(\gamma)$ (Eq.~(\ref{Rz})). 
Then, the total transformation matrix from $xyz$ crystal coordinate system to $x''y''z''$ setup coordinate system is 
\begin{equation}
\mathbf{R}^{(111)}=\mathbf{R}_z(\gamma)\mathbf{R}_{(-xy)}(\alpha)
\end{equation}
and we can transform elements of quadratic magneto-optical permittivity tensor similarly as in Eq.~(\ref{epnew})
\begin{equation}
\varepsilon^{(111)}_{ij}=R^{(111)}_{ik}R^{(111)}_{jl}
G_{klmn}{R^{(111)}}^{-1}_{ms}M''_s{R^{(111)}}^{-1}_{nt}M''_t
\,.
\label{epnew3one}
\end{equation}
The outgoing off-diagonal permittivity elements of $\varepsilon^{(111)}_{ij}$ and AMR-like parameter $\ep^{(111)}_{xx}-\ep^{(111)}_{yy}$ are presented in Tab.~\ref{t:111MMoff}, whereas 
diagonal elements of $\varepsilon^{(111)}_{ij}$ are presented in the Tab.~\ref{t:111MMdiag}.
Finally, Tab.~\ref{t:111Mpolar} presents elements of $\varepsilon^{(111)}_{ij}$ expressed as a function of the spherical coordinates of magnetization
vector ${\mathbf M}''$ (Eq.~(\ref{eq:Mdef})).

\begin{figure}[h]
\centering
\includegraphics[scale=0.5,angle=0]{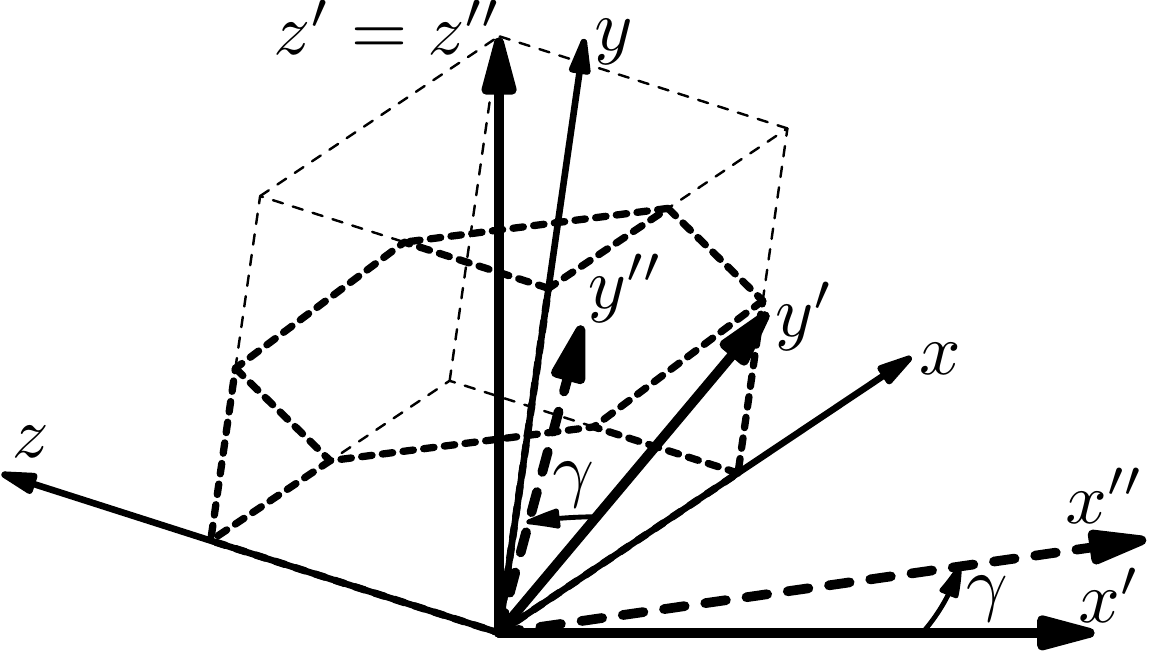}
\caption{\label{f:five}%
Transformation of coordinate system for sample with (111) oriented surface.}
\end{figure}

\begin{table*}
 \begin{tabular}{|>{\centering}m{0.115\textwidth}||>{\centering}m{0.40\textwidth}|>{\centering}m{0.40\textwidth}|}
 \hline
 & in-plane $\mathbf{M}$-scan ($\theta=\pi/2$), (001) oriented crystal
 & out-of-plane-II $\mathbf{M}$-scan, $\varphi=0$
 \tabularnewline\hline
  $\ds\ep_{yz}^{(001)}=\ep_{zy}^{(001)}$ 
  & $\ds 0$ 
  & $\ds 0$ 
 \tabularnewline\hline
  $\ds\ep_{xz}^{(001)}=\ep_{zx}^{(001)}$
  & $\ds 0$
  & $\ds G_{44} \sin 2\theta$
 \tabularnewline\hline
  $\ds \ep_{xy}^{(001)}=\ep_{yx}^{(001)}$
  & $\ds G_{44}\sin2\varphi-\frac{1}{2}\Delta G\sin2\gamma\cos(2\gamma+2\varphi) 
  \newline 
  - \frac{1}{2}\Delta\Gamma\sin2\gamma$
  & $\ds -\frac{1}{4}\Delta G\sin4\gamma\sin^2\theta 
  \newline
  + \frac{1}{2}\Delta\Gamma\sin2\gamma(2-3\sin^2\theta)$
 \tabularnewline\hline
  $\ds \ep_{xx}^{(001)}$
  & $\ds G_{11}-2G_{44}\sin^2\varphi+\frac{1}{2}\Delta G[\cos2\gamma\cos(2\gamma+2\varphi)-1] 
    \newline
    +\Delta\Gamma \sin\varphi\sin(2\gamma+\varphi)$ 
  & $\ds G_{11} - 2 G_{44}\cos^2\theta +\frac{1}{4}\Delta G[\sin^2\theta(3 +\cos4\gamma)-4] 
  \newline
  -\Delta\Gamma \cos2\gamma \cos^2\theta$    
 \tabularnewline\hline
  $\ds \ep_{yy}^{(001)}$
  & $\ds G_{11} - 2G_{44}\cos^2\varphi -\frac{1}{2}\Delta G[\cos2\gamma\cos(2\gamma+2\varphi)+1] 
    \newline
    - \Delta\Gamma\cos\varphi\cos(2\gamma+\varphi)$
  & $\ds G_{11} -2G_{44} +\frac{1}{4}\Delta G [\sin^2\theta(1-\cos4\gamma) -4] 
  \newline
  +\Delta\Gamma\cos2\gamma\cos2\theta$    
 \tabularnewline\hline
  $\ds \ep_{zz}^{(001)}$
  & $\ds G_{11}-2G_{44}-\Delta G +\Delta\Gamma \cos(2\gamma+2\varphi)$
  & $\ds G_{11} -2G_{44}\sin^2\theta -\Delta G \sin^2\theta +\Delta\Gamma \sin^2\theta\cos2\gamma$  
 \tabularnewline\hline
  $\ds \ep_{xx}^{(001)}-\ep_{yy}^{(001)}$
  & $\ds 2G_{44}\cos2\varphi + \Delta G \cos2\gamma\cos(2\gamma+2\varphi) 
  \newline
  +\Delta\Gamma \cos2\gamma$
  & $\ds 2G_{44}\sin^2\theta +\frac{1}{2}\Delta G\sin^2\theta(1+\cos4\gamma) 
    \newline
    +\Delta\Gamma\cos2\gamma(3\sin^2\theta-2)$  
  \tabularnewline\hline 
 \end{tabular}
 
 \begin{tabular}{|>{\centering}m{0.115\textwidth}||>{\centering}m{0.826\textwidth}|}
 \hline
 &  out-of-plane-I $\mathbf{M}$-scan ($\varphi=\pi/4$), (001) oriented crystal 
 \tabularnewline\hline
  $\ds\ep_{yz}^{(001)}=\ep_{zy}^{(001)}$ 
  & $\ds \frac{1}{\sqrt{2}} G_{44}\sin2\theta$ 
  \tabularnewline\hline
  $\ds\ep_{xz}^{(001)}=\ep_{zx}^{(001)}$
  & $\ds\frac{1}{\sqrt{2}}G_{44}\sin2\theta$ 
  \tabularnewline\hline
  $\ds \ep_{xy}^{(001)}=\ep_{yx}^{(001)}$
  & $\ds G_{44}\sin^2\theta + \frac{1}{2}\Delta G\sin^2\theta\sin^22\gamma 
    + \frac{1}{2}\Delta\Gamma\sin2\gamma(2-3\sin^2\theta)$
  \tabularnewline\hline
  $\ep_{xx}^{(001)}$
  & $\ds G_{11} - 2G_{44}(1-\frac{1}{2}\sin^2\theta) +\frac{1}{4}\Delta G[\sin^2\theta(2-\sin4\gamma)-4] 
  +\frac{1}{2}\Delta\Gamma[\sin^2\theta(\sin2\gamma+3\cos2\gamma)-2\cos2\gamma]
  $
  \tabularnewline\hline
  $\ds \ep_{yy}^{(001)}$
  & $\ds G_{11} - 2G_{44}(1-\frac{1}{2}\sin^2\theta) +\frac{1}{4}\Delta G[\sin^2\theta(2+\sin4\gamma)-4] 
  +\frac{1}{2}\Delta\Gamma[\sin^2\theta(\sin2\gamma-3\cos2\gamma)+2\cos2\gamma]
  $
  \tabularnewline\hline
  $\ds \ep_{zz}^{(001)}$
  & $\ds G_{11} -2G_{44}\sin^2\theta -\Delta G \sin^2\theta -\Delta\Gamma \sin2\gamma\sin^2\theta$
  \tabularnewline\hline
  $\ds \ep_{xx}^{(001)}-\ep_{yy}^{(001)}$
  & $\ds -\frac{1}{2}\Delta G \sin^2\theta\sin4\gamma +\Delta\Gamma \cos2\gamma(3\sin^2\theta-2)$   
  \tabularnewline\hline
 \end{tabular}
  \caption{\label{t:scan001}%
  Permittivity elements on magnetization orientation $\theta$, $\varphi$ for various magnetization scans for (001) oriented surface.}
\end{table*}

\begin{table*}
 \begin{tabular}{|>{\centering}m{0.07\textwidth}||>{\centering}m{0.423\textwidth}|>{\centering}m{0.423\textwidth}|}
\hline
 & in-plane $\mathbf{M}$-scan ($\theta=\pi/2$), (011) oriented crystal
 & in-plane $\mathbf{M}$-scan ($\theta=\pi/2$), (111) oriented crystal
  \tabularnewline\hline
 $\ds\ep_{yz}=\ep_{zy}$ 
 & $\ds -\frac{1}{4}\Delta\Gamma\cos\gamma[1+3\cos(2\gamma+2\varphi)]$
 & $\ds -\frac{1}{6}\Delta G [\cos(3\gamma+2\varphi)+\sin(3\gamma+2\varphi)]
  \newline
 -\frac{\sqrt{3}}{6}\Delta\Gamma[\cos(3\gamma+2\varphi)-\sin(3\gamma+2\varphi)]$
  \tabularnewline\hline
 $\ds\ep_{xz}=\ep_{zx}$ 
 & $\ds-\frac{1}{4}\Delta\Gamma\sin\gamma[1+3\cos(2\gamma+2\varphi)]$
 & $\ds \frac{1}{6}\Delta G [\cos(3\gamma+2\varphi)-\sin(3\gamma+2\varphi)]
  \newline
 -\frac{\sqrt{3}}{6}\Delta\Gamma[\cos(3\gamma+2\varphi)+\sin(3\gamma+2\varphi)]$
  \tabularnewline\hline
  $\ds\ep_{xy}=\ep_{yx}$ 
  & $\ds G_{44}\sin2\varphi - \frac{1}{8}\Delta G\sin2\gamma [1+3\cos(2\gamma+2\varphi)] $
& $\ds G_{44}\sin2\varphi +\frac{1}{6}\Delta G\sin2\varphi
  +\frac{\sqrt{3}}{6}\Delta\Gamma\cos2\varphi$
  \tabularnewline\hline
  $\ds\ep_{xx}$
  & $\ds G_{11} - 2G_{44}\sin^2\varphi 
  \newline
  + \frac{1}{8}\Delta G[ \cos(2\gamma+2\varphi)(1+3\cos2\gamma) +\cos2\gamma -5 ]$
 & $\ds G_{11}-2G_{44}\sin^2\varphi+\frac{1}{6}\Delta G(\cos2\varphi-4)
    -\frac{\sqrt{3}}{6}\Delta \Gamma\sin2\varphi$   
  \tabularnewline\hline
   $\ds\ep_{yy}$
  & $\ds G_{11} - 2G_{44}\cos^2\varphi 
  \newline
  + \frac{1}{8}\Delta G[\cos(2\gamma+2\varphi) (1-3\cos2\gamma) -\cos2\gamma -5 ]$ 
 & $\ds G_{11}-2G_{44}\cos^2\varphi-\frac{1}{6}\Delta G(\cos2\varphi+4)
    +\frac{\sqrt{3}}{6}\Delta \Gamma\sin2\varphi$   
  \tabularnewline\hline
  $\ds \ep_{zz}$
  & $\ds G_{11}-2G_{44}-\frac{1}{4}\Delta G[\cos(2\gamma+2\varphi)+3]$
 & $\ds G_{11}-2G_{44}-\frac{2}{3}\Delta G$   
  \tabularnewline\hline
  $\ds \ep_{xx}-\ep_{yy}$
  & $\ds 2G_{44}\cos2\varphi + \frac{1}{4}\Delta G\cos2\gamma[3\cos(2\gamma+2\varphi)+1]$  
 & $\ds 2G_{44}\cos2\varphi+\frac{1}{3}\Delta G\cos2\varphi
-\frac{\sqrt{3}}{3}\Delta \Gamma\sin2\varphi$   
  \tabularnewline\hline
\end{tabular}
\caption{\label{t:inplane011}\label{t:inplane111}
Permittivity elements for in-plane magnetization scan in case of (011) and (111) oriented surface.}
\end{table*}

\section{Discussion}

First let us discuss possibility to separate $G_{ij}$ quadratic elements from hypothetical experimental data or ab-initio calculations.

To determine experimentally quadratic elements, $G_{ij}$, in case of optical investigations, one can vary field direction (described by angles $\varphi$ and $\theta$) or one can rotate sample around its normal axis (described by an angle $\gamma$). In case of the conductivity measurements, the equivalent of the sample rotation corresponds to change of the direction of the applied probing in-plane current with respect to crystal axis. 
Namely, it is equivalent to the case when current flows in a fixed direction with respect to the setup cartesian system, and only crystal itself is rotated. Therefore, from symmetry point of view, following discussion is valid for both optical and transport investigations.

To separate quadratic terms $G_{ij}$, we discuss hypothetical measurements or ab-initio calculations for three magnetization scans (Tab.~\ref{t:scan001}). The first scan, ``in-plane magnetization scan'', keeps magnetization within $x''y''$-plane, i.e.\ the magnetization orientation is described as $\theta=\pi/2$ and $\varphi$ varies. 
This scan is particularly easy to obtain by experimental means, as well as (for $\gamma=0$) by ab-initio calculations. Second magnetization scan, ``out-of-plane-I magnetization scan'' goes from $z''$-direction
(i.e.\ from out-of-plane magnetization, i.e.\ from $\theta=0$) through $(111)''$ direction towards $(110)''$. This scan corresponds to $\varphi=\pi/4$ with varying $\theta$. 
This scan is particularly important for ab-initio calculations because in case of (001) oriented crystal with $\gamma=0$,  it passes all three types of high symmetry points, (001), (111) and (110), providing tough test of ab-initio calculations \cite{legut-private}.  The last magnetization scan to be discuss is called ``out-of-plane-II magnetization scan'' being defined as $\varphi=0$ and $\theta$ varies, i.e.\ it goes from out-of-plane magnetization orientation $(001)''$ to in-plane orientation $(100)''$.

\subsection{Magnetization scans for (001) oriented surface} 
First we treats in detail (001) oriented sample. The dependence of all permittivity elements for each magnetization scan is presented in Tab.~\ref{t:scan001}. 
(i) It reveals that the term $G_{11}$ is not accessible by any means, as it stays constant for any sample orientation and magnetic field orientation.  Hence it can not be extracted. 
(ii) The in-plane scan can provide values of all $G_{44}$, $\Delta G$ and $\Delta\Gamma$ from both the off-diagonal element (Hall-like) investigations and diagonal element (AMR-like) investigations. The extraction is based on different periodicity of the quadratic elements on $\gamma$. 
However, in case of $\Delta\Gamma$, this may be difficult to extract $\Delta\Gamma\sim \sin2\gamma$, as experimental artefacts (such as misaligned sample) may have the same symmetry. 
(iii) The most reliable way for extraction of $\Delta\Gamma$ is from $\ep_{zz}^{(001)}$, as it varies solely on $\Delta\Gamma$ through parameters $\varphi$ and $\gamma$. However, from experimental point of view, this element is very difficult to obtain by optical investigations and in case of the conductivity investigations, it requires current-perpendicular-to-plane geometry. 
(iv) In case of out-of-plane magnetization ($\theta=0$), the off-diagonal element $\ep_{xy}^{(001)}$ simplifies to $\ep_{xy}^{(001)}=\Delta\Gamma\sin2\gamma$. 
This provides probably the simplest way to determine $\Delta\Gamma$ by optical investigations, simply by observing anisotropy of the polar ($\theta=0$) Kerr effect (or Hall effect or XMCD) on sample orientation $\gamma$. 
(v) Finally, Table~\ref{t:scan001} demonstrates, that in case of both out-of-plane scans, the periodicity of different permittivity elements on $\gamma$ is similar as in case of in-plane magnetization scan. Also, no simple separation of quadratic term appears. Therefore, to separate quadratic terms $G_{ij}$, there is no advantage to apply magnetization scans going from in-plane to out-of-plane directions.

\subsection{In-plane magnetization scan for (011) and (111) oriented surface}

In previous paragraph, we have demonstrated that there is no particular advantage from experimental point of view to use out-of-plane magnetization scans to separate contributions from different quadratic elements for (001) surface oriented crystals. It can be shown that the same is valid for (011) and (111) oriented crystals. Therefore, in following we focus on dependence of permittivity elements only for in-plane magnetization scan, presented in Tabs.~\ref{t:inplane011}, respectively. 

In case of (011) surface probed with the in-plane magnetization scan (Tab.~\ref{t:inplane011}, left column),  
(i) both Hall-like term ($\ep^{(011)}_{xy}$) and AMR-like term ($\ep_{xx}^{(001)}-\ep_{yy}^{(001)}$)  depend solely on $G_{44}$ and $\Delta G$ and their value can be extracted by both varying magnetic field $\varphi$ or sample orientation $\gamma$. Like in (001) surface oriented crystal, $G_{44}$ contribution is independent on sample orientation $\gamma$ whereas $\Delta G$ contribution varies with $\gamma$. Interestingly, in later case the dependence of $\Delta G$ has periodicity on both $2\gamma$ and $4\gamma$. (ii) $\Delta\Gamma$ is accessible solely through $\ep_{yz}^{(011)}$ and $\ep_{xz}^{(011)}$ terms. However, when magnetization goes out-of-plane, dependence of all terms on $\Delta\Gamma$ appear (see Tab.~\ref{t:111MMdiag}).
 
In case of (111) oriented surface  probed with in-plane magnetization scan (Tab.~\ref{t:inplane011}, right column), remarkable point is that the periodicity is either $2\varphi$ (when rotating magnetization) or $3\gamma$ (when rotating sample).
For $\ep_{xy}^{(111)}$ (Hall-like), $\ep_{xx}^{(111)}-\ep_{yy}^{(111)}$ (AMR-like) permittivity elements and for all diagonal permittivity elements, all three quadratic elements $G_{44}$, $\Delta G$ and $\Delta\Gamma$ contributions are either constants or they depend on magnetization orientation as $2\varphi$, all of them being independent on $\gamma$. Remaining permittivity elements $\ep_{xz}^{(111)}$ and $\ep_{yz}^{(111)}$ depends only on $\Delta G$ and $\Delta\Gamma$ and both contributions depend in a similar  on $\gamma$ and $\varphi$. Therefore, separation of quadratic elements from (111) oriented crystal is shown to be particularly difficult.

\section{Conclusion}
Using symmetry arguments, we determined diagonal and off-diagonal elements of the permittivity (or conductivity) tensor up to the second order in magnetization for cubic crystals. We express all tensor elements for arbitrary magnetization direction and for (001), (011) and (111) surface orientations. Finally, we discuss a way to separate various quadratic elements of the second order permittivity tensors for different magnetization direction and different surface orientations in case of both optical and transport investigations.

\begin{acknowledgement}
Financial support by Structural Funds of the European Union and state budget of the Czech Republic (Nanobase project CZ.1.07/2.3.00/20.0074) and by the IT4Inno\-va\-tions Centre of Excellence project, CZ.1.05/1.1.00/02.0070, as well as fruitful discussions with J.~Vl\v{c}ek are well acknowledged.
\end{acknowledgement}

\appendix
\section{Linear-in-magnetization permittivity and conductivity tensor for cubic crystals}

Although this Article treats permittivity/conductivity tensor in the second order in magnetization, let us shortly overview their dependence on the zeroth and the first order in magnetization. 

In case of the zeroth-order permittivity $\ep^{(0)}_{ij}$ in the cubic crystal, the permittivity is independent on both crystal and field orientation, $\ep^{(0)}_{ij}=\ep^{(0)} \delta_{ij}$, where $\delta_{ij}$  is Kronecker delta, and $\ep^{(0)}$ is scalar zero-order permittivity.

In case of permittivity elements being in the first order in magnetization, $\ep^{(1)}_{ij}$, their dependence is summarized in Tab.~\ref{t:firstorder}, written in similar notation as all other Tables introducing the second-order permittivity. Note that $\ep_{ij}^{(1)}$ does not depend on crystal orientation, so Tab.~\ref{t:firstorder} is valid for any surface orientation of the crystal and any sample orientation $\gamma$.  
\begin{table}
  \begin{tabular}{|c||c|c|c|}
    \hline
    & $\ds M_x=$ & $\ds M_y=$ & $\ds M_z=$
    \\ 
    & $\ds \sin\theta\cos\varphi$ & $\ds \sin\theta\sin\varphi$ & $\ds \cos\theta$
    \\ \hline
    $\ds\ep^{(1)}_{yz}=-\ep_{zy}^{(1)}$ & $K $ & $0$ & $0$
    \\ \hline
    $\ds\ep^{(1)}_{zx}=-\ep_{xz}^{(1)}$ & $0$ & $K$ & $0$
    \\ \hline
    $\ds\ep^{(1)}_{xy}=-\ep_{yx}^{(1)}$ & $0$ & $0$ & $K$
    \\ \hline
  \end{tabular}
  \caption{The first-order-in-magnetization permittivity elements $\ep_{ij}^{(1)}$ as a function of magnetic field orientation for cubic crystal. $\ep_{ij}^{(1)}$ does not depend on crystal orientation.}
\label{t:firstorder}
\end{table}


\providecommand{\WileyBibTextsc}{}
\let\textsc\WileyBibTextsc
\providecommand{\othercit}{}
\providecommand{\jr}[1]{#1}
\providecommand{\etal}{~et~al.}

\end{document}